\documentclass[conference]{IEEEtran}
\ifCLASSINFOpdf
\usepackage[pdftex]{graphicx}
\graphicspath{{../pics}} \DeclareGraphicsExtensions{.eps} \else
\usepackage[dvips]{graphicx}
\graphicspath{{../pics}} \DeclareGraphicsExtensions{.eps} \fi
\usepackage{kbordermatrix}

\begin{document}

\title{Visualizing Cognitive Moves for Assessing Information Perception Biases in Decision Making}

\author{

\IEEEauthorblockN{Antony W. Iorio}

\IEEEauthorblockA{University of New South Wales, ADFA, Canberra ACT 2600, Australia.
Antony.Iorio@gmail.com}

\IEEEauthorblockN{Hussein A. Abbass}

\IEEEauthorblockA{University of New South Wales, ADFA, Canberra ACT 2600, Australia.
h.abbass@adfa.edu.au}

\IEEEauthorblockN{Svetoslav Gaidow }

\IEEEauthorblockA{Defence Science and Technology Organisation, Edinburgh SA 5111, Australia.
Svetoslav.Gaidow@dsto.defence.gov.au}

\IEEEauthorblockN{Axel Bender}

\IEEEauthorblockA{Defence Science and Technology Organisation, Edinburgh SA 5111, Australia.
Axel.Bender@dsto.defence.gov.au}

} \maketitle

\begin{abstract}
In decision making a key source of uncertainty is people's perception of information which is
influenced by their attitudes toward risk. Both, perception of information and risk attitude,
affect the interpretation of information and hence the choice of suitable courses of action in a
variety of contexts ranging from project planning to military operations. Visualization associated
with the dynamics of cognitive states of people processing information and making decision is
therefore not only important for analysis but has also significant practical applications, in
particular in the military command and control domain. In this paper, we focus on a major concept
that affect human cognition in this context: reliability of information. We introduce Cognitive
Move Diagrams (CMD)---a simple visualization tool---to represent and evaluate the impact of this
concept on decision making. We demonstrate through both a hypothetical example and a subject matter
expert based experiment that CMD are effective in visualizing, detecting and qualifying human
biases.
\end{abstract}

\IEEEpeerreviewmaketitle

\section{Introduction}\label{sec:Introduction}
The field of cognition is concerned with mental contents and the processes by which these contents
are manipulated. A challenge facing the cognitive sciences is that most mental contents and
associated mental processes are invisible. Attempts are thus being made to illuminate this black
box of human mental processes by way of statistical inferencing supported by data from
neuro-physiological monitoring.

Visualization is normally used to represent neuro-phy\-si\-o\-lo\-gi\-cal data such as signals of
brain activity; data generated from statistical analysis such as a scatter diagram with a line fit
of some stimuli categories and the selected action; or participant data such as a graph of how a
group of children organizes a set of concepts. However, visualization has a lot more to offer to
the cognitive sciences. Simple visualization tools can provide powerful, self-explanatory messages
during an analysis phase. In this paper, our aim is to visually represent dynamic cognitive
elements, contents, and/or processes of a group. We hence develop a type of diagrammatic
representation that we call Cognitive Move Diagrams (CMD).

\section{Cognitive Move Diagrams}\label{sec:CMD}
Situational awareness is a cognitive process that characterizes how, in a conscious manner, an
agent translates input information to an internal representation of context. To facilitate the
transmission of complex high-dimensional information, e.g. in a military battle space, one can use
a Common Operating Picture (COP). Here, myriads of pieces of sensor data are fused and get
represented in a richer language than that used to describe the raw information. This language can
be a visual representation of the situation, e.g.\ a map with a heat diagram of activities; a
graphical network with arcs representing recorded communication among nodes (entities); or a
higher-order logic or a system of equations that derive from the fused sensory information the
trajectories of a group of unmanned aerial vehicles. In short, a COP compresses the sensory
information to a representation of richer semantics that can easily be communicated to a decision
maker. CMD are in effect COP of the dynamics of mental contents in a group. They facilitate the
understanding of how small changes---such as introducing a new concept to a
scenario/situation---affect consent or dissent of a group with respect to the interpretation of the
situation. In some sense, CMD is a graphical representation of the dynamics of the mental content
of a group as a function of physical and/or psychological stimuli/affects. However, as will be
illustrated in the remainder of this paper, CMD can have a much wider application than that.

As mentioned above, researchers need to measure the inaccessible mental processes of interest
through some proxies or a set of dependent/outcome variables of an experiment or a natural
situation. Let us define the vector~$\vec{C}^t = [\vec{c^t}_1, \dots, \vec{c^t}_N]$ of outcome
variables (vectors) about which we collected some measurements~$\vec{V}^t$ at time~$t$. These
variables are proxies for the concepts of interest. $\vec{c^t}_i, \ \ i=1,\dots,N$ is the vector of
outcome~$i$ and has dimension~$M$, with $M$ the number of agents/humans. In other words, for each
outcome variable, we have readings for all $M$ agents. $\vec{V}$ is a matrix of measurements and
has a dimension of $M \times K$, with $K$ the number of independent variables used to manipulate
the experimental setup. Manipulation occurs through varying the experimental conditions ({\it i.e.}
scenarios) at different time steps~$t=1,\dots,T$, with $T$ representing the number of times the
scenarios change in the whole experiment. The independent variables represent the stimuli used to
drive cognitive dynamics in the $M$ different agents.

We now introduce the following definitions:
\begin{itemize}
\item {\bf Agent Cognitive State (ACS)} The cognitive state $\vec{s}^t_j$ for agent~$j$ at time~$t$
is defined by the outcome variable measurement vector $\vec{s}^t_j = [ c^t_{1j}, \dots, c^t_{Nj}]$.
\item {\bf Agent Cognitive Move (ACM)} When the measurement vector for agent~$j$ changes because of
a manipulation of the scenario at time step~$t$, the transition from the old ACS, $\vec{s}^t_j$, to
the new ACS, $\vec{s}^{t+1}_j$, is an ACM $\Delta\vec{s}^{t}_j$.
\item {\bf Group Cognitive State (GCS)} The cognitive state~$\vec{S}^t_j$ for a group of agents
$j=1,\dots,M$ at time $t$ is defined by a functional mapping from ACS to a matrix of similar
dimension, $ GCS: [ \vec{s}^t_j \rightarrow \vec{S}^t_j ]$. An example of this mapping can simply
be a function that clusters the agents together and replaces the value of each outcome variable
with a cluster variable (e.g.\ centroid). Therefore, agents falling in the same cluster at time~$t$
will have identical values; i.e., $S^t_{ij} = S^t_{il} $ if and only if both $j$ and $l$ fall in
the same cluster; that is, $similar(j,l)=yes$, where $similar$ is the similarity function used for
the clustering.
\item {\bf Group Cognitive Move (GCM)}  $GCM\,\Delta\vec{S}^{t}: \vec{S}^t \rightarrow S^{t+1}$;
thus a GCM~$\Delta\vec{S}^{t}$ is the transition from the GCS at one time step to the next time
step.
\item A {\bf Cognitive Move Diagram (CMD)} is a representation of the state transition diagram of
GCM using some encoding scheme.
\end{itemize}

\section{Example}

We illustrate the previous definitions in a hypothetical example. This example deviates from our
primary use of CMD, the visualization of human biases related to information reliability and
temporal dependence of decision, so as to demonstrate that the visualization can be generalized and
of benefit to the wider cognitive sciences literature.

In a military context, imagine we are interested to understand the impact of weapons noise on
memory. Let's assume $M=3$ individuals participated in a within-group experiment. Each experiment
involved memory testing that was performed on three days, each separated by a week. For each
individual, the sequence in which the experimental conditions varied was chosen randomly. Twenty
words and their meanings were presented to the subjects in a random order. Two hours later, they
were first asked to recall as many words as they could. They were then again presented with the
same twenty words and were asked to cluster them according to meaning. The conditions on the test
days varied such that on one of the three days, the individuals were not exposed to weapons noise
within these two hours, on the other two days, the subject was exposed to a one-hour and two-hour
weapons noise during the two hours between being told the words and being tested.

In this experiment, there is $K=1$ ordinal independent variable---the length of time during which a
subject was exposed to the sound of a gun fire, i.e. $t\in\{0,1,2\}$~(hours). $N=2$ dependent
variables were ratio measured: the percentage of words recalled and the percentage of successful
associations of words with concepts.

\subsection{Outcome Variables and Measurements:}

$\vec{C}^0 = [[0.20 \ 0.25 \ 0.45] [0.50 \ 0.50 \ 0.70]]$

$\vec{C}^1 = [[0.70 \ 0.75 \ 0.70] [0.70 \ 0.70 \ 0.75]]$

$\vec{C}^2 = [[0.25 \ 0.50 \ 0.75] [1.00 \ 0.60 \ 0.55]]$

$\vec{V^0} = [0 \ 1 \ 2]$; \ \ $\vec{V^1} = [0 \ 1 \ 2]$; \ \ $\vec{V^2} = [0 \ 1 \ 2]$

\subsection{Agent Cognitive States:}

$\vec{s}^0_1 = [0.20 \ 0.50]; \ \ \vec{s}^0_2 = [0.25 \ 0.50]; \ \ \vec{s}^0_3 = [0.45 \ 0.70]$

$\vec{s}^1_1 = [0.70 \ 0.70]; \ \ \vec{s}^1_2 = [0.75 \ 0.70]; \ \ \vec{s}^1_3 = [0.70 \ 0.75]$

$\vec{s}^2_1 = [0.25 \ 1.00]; \ \ \vec{s}^2_2 = [0.50 \ 0.60]; \ \ \vec{s}^2_3 = [0.75 \ 0.55]$

\subsection{Agent Cognitive Moves:}
\begin{eqnarray*}
\mbox{Agent 1} & = & [0.20 \ 0.50] \rightarrow [0.70 \ 0.70] \; \mbox{and}\\
               &   & [0.70 \ 0.70] \rightarrow [0.25 \ 1.00]\\
\mbox{Agent 2} & = & [0.25 \ 0.50] \rightarrow [0.75 \ 0.70] \; \mbox{and}\\
               &   & [0.75 \ 0.70] \rightarrow [0.50 \ 0.60]\\
\mbox{Agent 3} & = & [0.45 \ 0.70] \rightarrow [0.70 \ 0.75] \; \mbox{and}\\
               &   & [0.70 \ 0.75] \rightarrow [0.75 \ 0.55]
\end{eqnarray*}

\subsection{Group Cognitive Moves:}

Assume we used some clustering algorithm for each weapons noise exposure time interval. The
algorithm for $t=0$ assigned the first two agents to the same cluster with a centroid of $[0.23 \
0.50]$ and the third agent to one cluster described by its original vector. For $t=1$ all agents
were in one and the same cluster with a centroid of $[0.72 \ 0.72]$. For the last time interval,
each agent occupied a different cluster. Therefore, the GCS is as follows:

$S^0_1 = [0.23 \ 0.50]; \ \ S^0_2 = [0.23 \ 0.50]; \ \ S^0_3 = [0.45 \ 0.70]$

$S^1_1 = [0.72 \ 0.72]; \ \ S^1_2 = [0.72 \ 0.72]; \ \ S^1_3 = [0.72 \ 0.72]$

$S^2_1 = [0.25 \ 1.00]; \ \ S^2_2 = [0.50 \ 0.60]; \ \ S^2_3 = [0.75 \ 0.55]$

The clustering algorithm simply revealed that from a group (similarity measure) perspective all
agents were in approximately the same state when exposed to gun fire noise for an hour. The GCM is
as follows:

\begin{eqnarray*}
\mbox{Agent 1} & = & [0.23 \ 0.50] \rightarrow [0.72 \ 0.72] \; \mbox{and}\\
               &   & [0.72 \ 0.72] \rightarrow [0.25 \ 1.00]\\
\mbox{Agent 2} & = & [0.23 \ 0.50] \rightarrow [0.72 \ 0.72] \; \mbox{and}\\
               &   & [0.72 \ 0.72] \rightarrow [0.50 \ 0.60]\\
\mbox{Agent 3} & = & [0.45 \ 0.70] \rightarrow [0.72 \ 0.72] \; \mbox{and}\\
               &   & [0.72 \ 0.72] \rightarrow [0.75 \ 0.55]
\end{eqnarray*}

\section{Encoding Schemes}

\begin{figure}[!t]
\centering
\includegraphics[width=3.1in, height=2.0in]{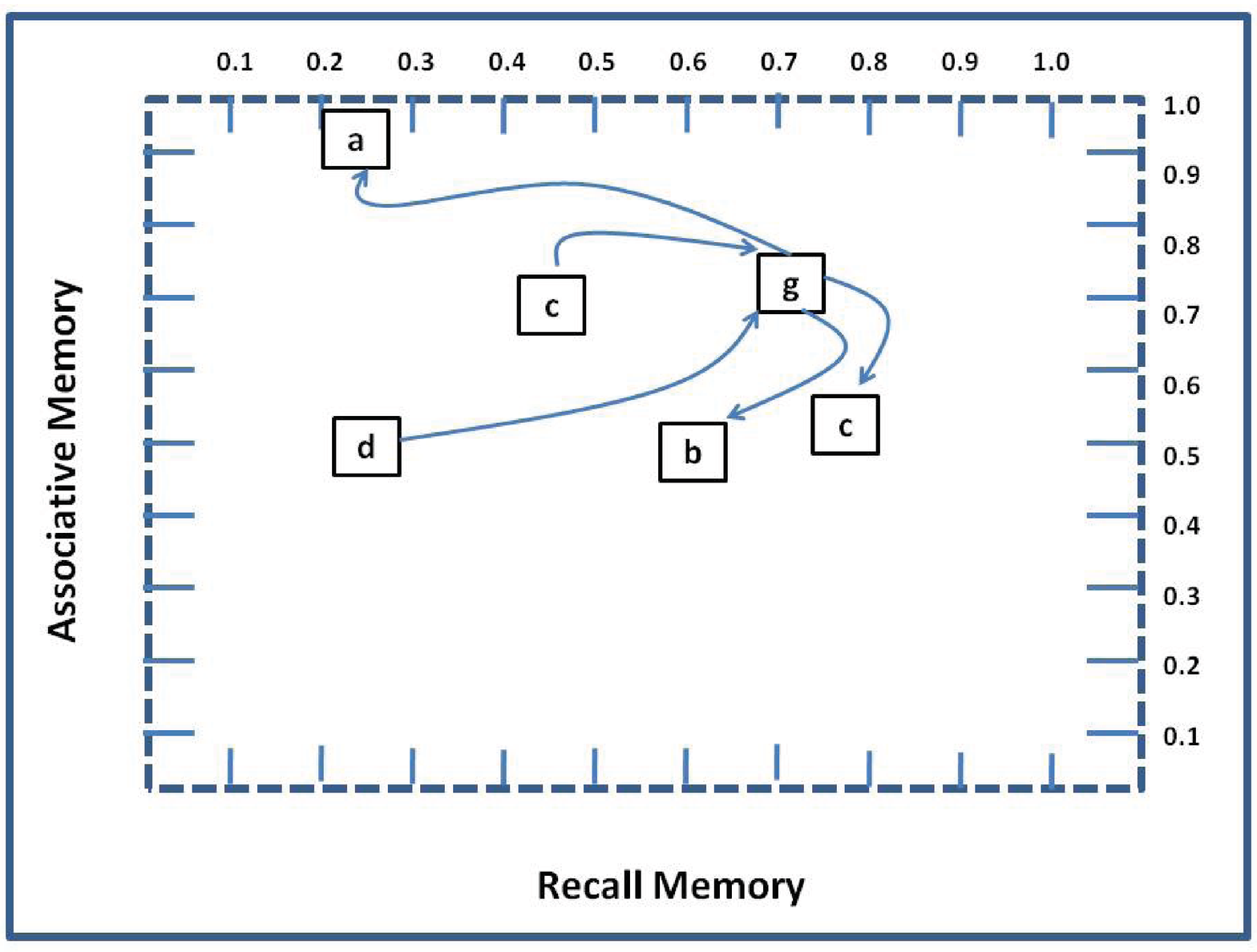}
\caption{CMD for the memory example using Encoding Scheme~1.} \label{fig:cmd1}
\end{figure}

We present three possible ways to represent GCM in a CMD. The first encoding scheme relies on
associating a unique symbol to each cluster. In our hypothetical example, there are seven possible
unique clusters. Thus, we associate an alphabetic letter to each cluster according to the following
rule: a cluster that contains only one of the three individuals is marked by letter `a', `b', and
`c' for individual 1, 2, and 3, respectively. A cluster of two individuals is labelled `d', `e', or
`f' when containing individuals (1,2), (1,3), or (2,3), respectively. When all individuals fall
into a single cluster, it is assigned the letter `g'. The resulting CMD is depicted in
Figure~\ref{fig:cmd1}.

Encoding Scheme~1 has the advantage that participants are uniquely associated with clusters. In its
visualization, the arrows that link clusters for different noise exposure times offer both a
glimpse of the dynamics of the GCS and the start and end states. For instance, one can immediately
see that noise exposure over a period of one hour unified the group memory ability. One can also
make out clearly that the first two individuals, despite starting with and maintaining similar
memory ability, diverged dramatically when exposed to two hours of noise.

\begin{figure}[!t]
\centering
\includegraphics[width=3.1in, height=2.0in]{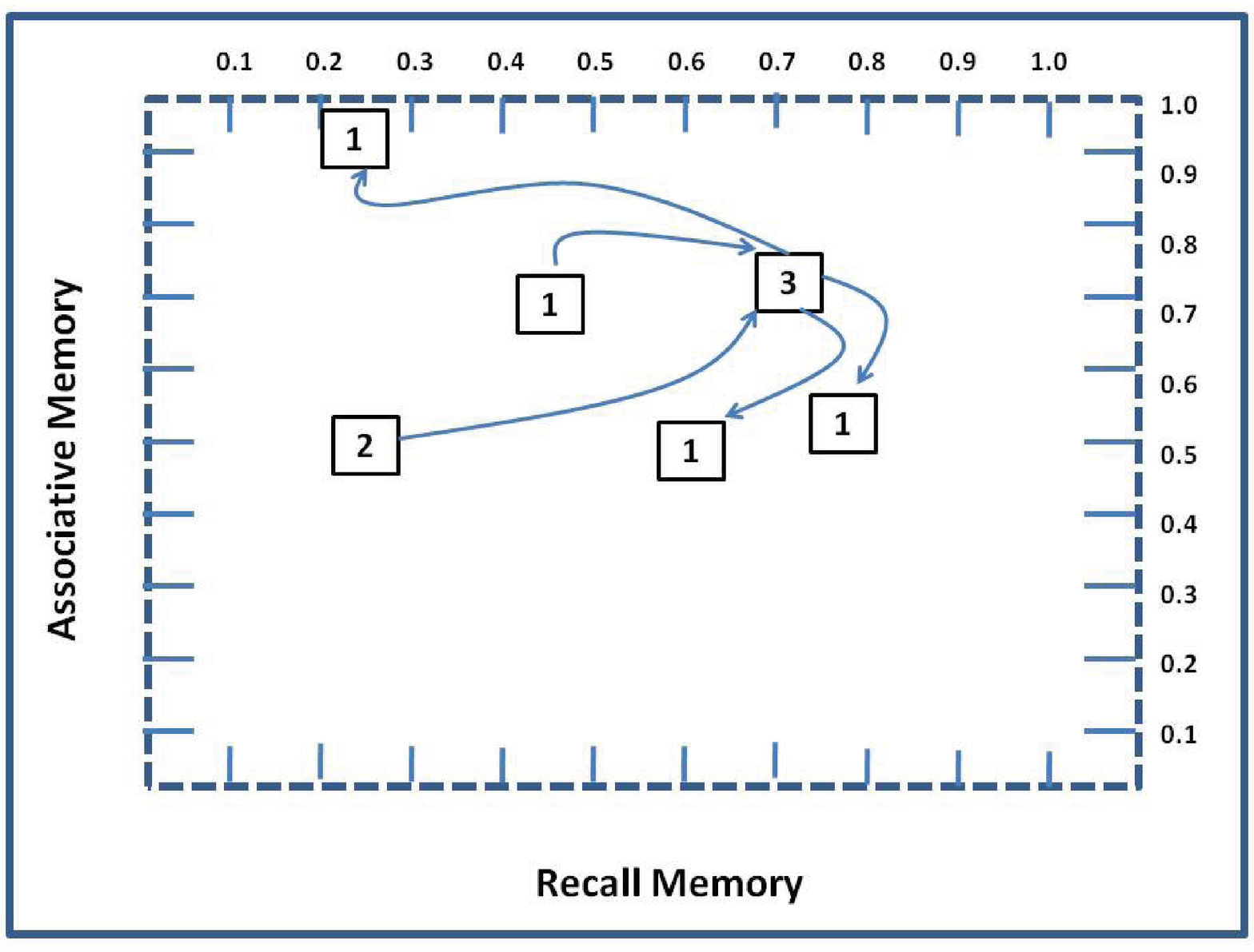}
\caption{CMD for the memory example using Encoding Scheme~2.} \label{fig:cmd2}
\end{figure}

The disadvantage of Encoding Scheme~1 is that it does not scale well as the number of individuals
increases. When numbers are large, it is better to label a cluster with the number of individuals
contained in it. While in this second encoding scheme information about individuals gets lost, this
information is typically not very useful anyway when cohorts are big. Figure~\ref{fig:cmd2} shows a
CMD with Encoding Scheme~2. As with Encoding Scheme~1 the cognitive changes of the group of
subjects exposed to different experimental conditions can be grasped quickly.

\begin{figure}[!t]
\centering
\includegraphics[width=3.1in, height=2.0in]{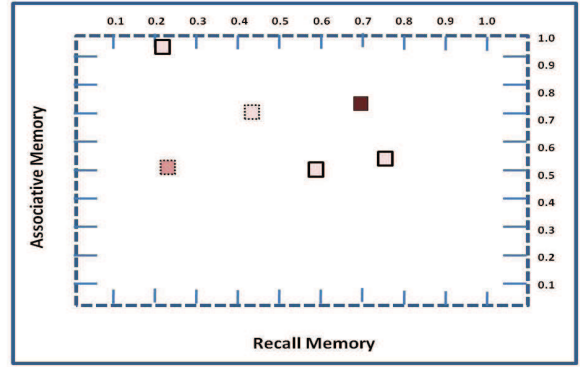}
\caption{CMD for the memory example using Encoding Scheme~3.} \label{fig:cmd3}
\end{figure}

If the number of GCM is very large, the transition lines will clutter the diagram. Our third
encoding scheme in Figure~\ref{fig:cmd3} uses colored boxes to represent clusters. Color intensity
represents cluster size: the darker the color, the greater the number of cluster members. The
boundary line of a box encodes the independent experimental variable: the thicker the line, the
larger is the independent variable.

\section{Scalability Issues}
The visualization of GCM in CMD is restricted to two or three dimensions. This is not necessarily a
limitation in experiments with small numbers of outcome variables. However, it can become
significant in our domain of application---military information processing---where a primary
objective is to have a generic situational awareness visualization method that depicts deviations
from expected cognitive moves in a group of soldiers. As such, often tens of factors need to be
analyzed simultaneously and displayed to project a group's behavior. In this case, our
recommendation is to group these factors and, instead of visualizing the raw outcomes, use
dimensionality reduction techniques such as principal component analysis. The visualization can
then be done, for instance, for the first two principal components only.

\section{Case Study}
In military command and control (C2), information reliability can have a sizeable impact on the
courses of action taken. Not knowing how reliable information is, equates to uncertainty. In the
context of decision making such uncertainty may affect the decision maker's objectives. How much a
decision is affected is likely to depend on his or her attitude to risk (the effect of uncertainty
on objectives).

Obviously, there exists a need to quantify how risk attitude affects decision making. The
literature is replete with methods, models, and studies for assessing risk attitude in humans. Risk
attitude has been assessed using games, such as the Angling Risk Task~\cite{art}, the Balloon
Analog Risk Task~\cite{bart}, the Columbia Card Task (CCT)~\cite{cct}, the Cups Task~\cite{cups},
and the Devil's Task (also known as Slovic's Risk Task)~\cite{devil}. However, these are very
specific games that situate players in specific, artificial environments. Risk attitude though
depends on context, in particular in military C2. Similarly, there are many measures of risk
attitude, such as the Attitudes to Risk Taking~\cite{art2}, Domain Specific Risk Attitude
(DoSpeRT)~\cite{dospert} and the Risk Propensity Scale also known as the Risk Taking
Index~\cite{rti}, etc. These measures are typically utilized in general risk assessments and are
rarely suited to military decision environments.

In a case study we posed the questions how much someone's decision changes when they are told that
the information presented to them was reliable? If the concept of information reliability did not
cross their mind, how does their behavior alter when they suddenly become aware of it? How much
does the framing of information reliability affect their decision making? We phrased these
questions in the context of the strategic planning of truck allocations to future operations.
Decision makers (``agents") were required to make their allocation decisions three times where the
only factor that changed from experiment to experiment was the framing of the information about the
reliability of truck demands. Specifically, in the first experiment scenarios were described
without making any reference to reliability of the provided demand interval estimates; in the
second experiment information about maximum and minimum demands across the various competing
scenarios was described qualitatively (e.g. ``very high"); in the third experiment, the reliability
of information was quantified. The provided reliability estimates constitute the manipulation
variable discussed in Section~\ref{sec:Introduction}. The experiment was set up such that resources
provided to participants were sufficient to meet the expected value of the demand intervals. The
AGS of the decision makers reflects their deviation from expected value in the context of
reliability information.

\begin{figure}[!t]
\centering
\includegraphics[width=3.1in, height=2.2in]{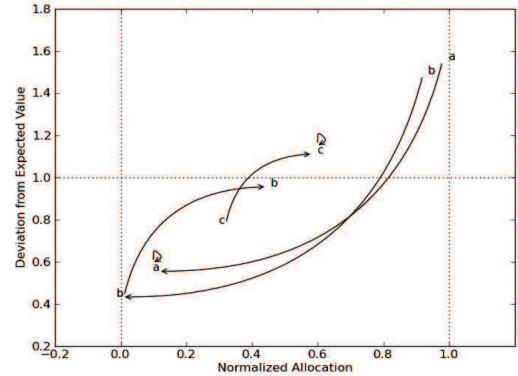}
\caption{CMD using Encoding Scheme~1 for the resource allocation problem presented to three
military subject matter experts.} \label{fig:mvcmexample}
\end{figure}

Figure~\ref{fig:mvcmexample} shows a CMD for three participants of the case study experiments with
the outcome measurement variables being the normalized truck allocations that participants made in
a given scenario and the deviation of their allocation from the expected value. The application of
clustering mappings resulted in clusters of single individuals only, labelled `a', `b' and `c' in
Figure~\ref{fig:mvcmexample}. From the CMD we can see at a glimpse that, firstly, the introduction
of the concept of Information Reliability had an impact on all the decisions made by the
participant. Secondly, the graph indicates that the group of participants does not have a shared
understanding of information reliability. For instance, the transition from State `a' to another
unique State `a' is not shared by any other participant.

\section{Conclusion}\label{sec:Conclusion}

In this paper we have introduced and developed Cognitive Move Diagrams (CMD), adapted from some
simple and general notions of individual and group cognitive states, and transitions between these
states. This technique for visualizing cognitive similarities and differences is robust for use in
a range of experiments with varying manipulation variables. We have presented an illustration of
CMD applied to a hypothetical case in which participants' associative memory might be affected by
gun fire noise. We also presented results from a real-world case study involving human
participants. Here, the effects of both the framing of information reliability and a decision's
time horizon were studied in the context of strategic defence procurement decision making. The
simple hypothetical example and the real-world case study demonstrate how decision makers who reach
conclusions independently of each other, move through a cognitive space. We illustrated how
similarities and differences in cognitive perception can be visualized in a CMD applying various
encoding schemes.

\bibliographystyle{IEEEtran}

\end{document}